\documentclass[12pt]{article}

\newbox\pmbbox
\def\pmb#1{{\setbox\pmbbox=\hbox{$#1$}%
   \copy\pmbbox \kern-\wd\pmbbox \kern.3pt \raise.3pt
    \copy\pmbbox \kern-\wd\pmbbox \kern.3pt\box\pmbbox}}

\newcommand{\be}{\hbox{\large e}}

\begin{document}

\title{Squeezing Operator and Squeeze Tomography}

\author{Octavio Casta\~nos, Ram\'on L\'opez-Pe\~na, \\
Margarita A. Man'ko\footnote{On leave from P.N.~Lebedev
Physical Institute, Moscow, Russia} ~and Vladimir I. Man'ko$^{\ast}$ \\
Instituto de Ciencias Nucleares, UNAM \\
AP~70-543, 04510 M\'exico, DF, M\'EXICO}

\date{}

\maketitle

\begin{abstract}
Some properties of Pleba\'nski squeezing operator and squeezed states created with time-dependent quadratic in position and momentum Hamiltonians are reviewed. New type of tomography of quantum states called squeeze tomography is discussed. 
\end{abstract}

\section{Introduction}

Last two decades the phenomenon of squeezing, especially squeezed states in quantum optics attracted a lot of attention (see, for example \cite{DMnewbook} where the review of nonclassical states of light is presented). An important contribution to the theory of squeezed states was made by Infeld and Pleba\'nski in studies \cite{Pleb54,InfPleb,Pleb55},
whose results were summarized in \cite{Pleb}. Pleba\'nski introduced the following family of states, described by the vector:
\begin{equation}
|\tilde\psi\rangle=\exp\left[i\left(\eta\hat{x}-\xi\hat{p}\right)\right]
\exp\left[\frac{i}{2}\log a\left(\hat{x}\hat{p}+\hat{p}\hat{x}
\right)\right] |\psi\rangle,
\label{Plebst}
\end{equation}
where $\hat x$ and $\hat p$ are position and momentum operators, respectively, $\xi$, $\eta$, and $a>0$ are real parameters, and $|\psi\rangle$ is an arbitrary initial state. Evidently, the first exponential in the right-hand side of (\ref{Plebst}) is the displacement operator written in terms of the Hermitian quadrature operators. Its properties were studied in \cite{Pleb54}. The second exponential is the special case of the squeezing operator (see Eq. (\ref{sq-op}) below).
 
For the initial vacuum state, $|\psi_0\rangle=|0\rangle$, the state $|\tilde\psi_0\rangle$ (\ref{Plebst}) is exactly the squeezed state in modern terminology, whereas choosing other initial states one can obtain various generalized squeezed states. In particular, the choice
$|\psi_n\rangle=|n\rangle$ results in the family of squeezed-number operator states, which were considered in \cite{Pleb55,Pleb}.

In the case $a=1$ (considered in \cite{Pleb54}), we arrive at the states known nowadays under the name ``displaced number states.''
Pleba\'nski gave the explicit expressions describing the time
evolution of the state (\ref{Plebst}) for the harmonic oscillator
with a constant frequency and proved the completeness of the set of
``displaced''-number operator states.

Infeld and Pleba\'nski \cite{InfPleb} performed a detailed study
of the properties of the unitary operator
$\exp(i\hat{T})$, where $\hat{T}$ is a generic inhomogeneous
quadratic form of the canonical operators $\hat{x}$ and $\hat{p}$
with constant $c$-number coefficients.

Stoler \cite{Stol1} showed that the minimum-uncertainty states
can be obtained from the oscillator ground state by means of the unitary operator depending on the complex number $z$, creation $\hat a^\dagger$ and annihilation $\hat a$ operators 
\begin{equation}
\hat{S}(z)=\exp\left[\frac{1}{2}\left(z\hat{a}^2 -z^* \hat{a}^{\dagger2}
\right)\right],
\label{sq-op}
\end{equation}
which was later on named the ``squeezed operator.'' In the second paper of \cite{Stol1} the operator $\hat{S}(z)$ was written for real $z$ in terms of the quadrature operators as
$\exp\left[ir\left(\hat{x}\hat{p}+\hat{p}\hat{x}\right)\right]$,
which is exactly the form given by Pleba\'nski \cite{Pleb}.
The conditions under which the minimum-uncertainty states preserve 
their form were studied, for example, in \cite{Trif74}.

Pleba\'nski squeezing operator can be also used to construct the specific scheme of measuring quantum states called squeeze tomography~\cite{JPAlast}. Below we review this scheme and give a short description of other tomography methods. One of possible methods to create squeezed states is using time-dependent Hamiltonians, e.g., Hamiltonian of parametric oscillator. The system with squeezing~\cite{Caldirola-Kanai} and quantum damping~\cite{elizabeth} is also described by such Hamiltonians. We will illustrate the squeezing phenomenon using the example of the damped oscillator.

The quantum states are described either by wave
functions~\cite{sch26} (pure states) or by density
matrix~\cite{landau27,vonneumann} (mixed states). The attempts to find a
description of the quantum states which more closely resembles to
the classical picture give rise to the quasidistribution functions
in phase space of field quadratures as Wigner
function~\cite{wig32}, Sudarshan--Glauber
$P$-function~\cite{sud63,gla63}, Husimi $Q$-function~\cite{hus40}.
Recently it was understood that the states can be associated with
the standard probability distribution functions. This
understanding emerged when the relation between the marginal
distribution function for photon homodyne quadrature (optical
tomogram) and Wigner function was found~\cite{ber-ber,vog-ris}.

Optical tomography of quantum states was used to measure the
quantum states of squeezed light~\cite{raym,mlyn}. The optical
tomograms depending on a rotation angle parameter were
generalized~\cite{tom95,d'ar96,wunsche} to the case of symplectic
tomograms of quantum states, which depend on a field quadrature
and two additional real parameters.
The symplectic and optical tomograms depend on random continuous
variables (homodyne quadrature components). There exists another
tomographic scheme to measure the quantum states. This scheme uses
probability distributions of discrete random variable $n=0, \, 1,
\, 2, \, \dots$, which has the physical meaning of number of
photons~\cite{wvogel,banaszeck,tombesi_epl}. 
 Another tomographic scheme is based on spin
tomography~\cite{dod,olga,klimov,castanos}, where discrete random variable is the spin projection $m$, $-j<m<j$.

The aim of our paper is to discuss the new tomographic
representation which we called squeeze tomography.
The squeeze tomogram uses the discrete random
variable $n=0, \, 1, \, 2, \, \dots$, which is the photon number
analogously to the case of photon-number tomography.

The examples of the squeeze tomograms for the coherent states~\cite{gla63}, even and odd coherent states~\cite{dmm74}, thermal states, and squeezed
states~\cite{vourd,schleich-wheeler} will be considered.

\section{Squezed states of parametric oscillator and Caldirola--Kanai Hamiltonians}

Squeezed states can be generated by the parametric excitation of oscillator.
In the case of parametric oscillator, the Hamiltonian is described by 
	 \begin{equation}
		H = \frac{1}{2} p^{2} + \frac{1}{2} \omega^{2} (t) \, q^{2} \ ,
	\end{equation}
where we take $\omega (0) = \hbar = m = 1$. There exist the time-dependent constants of
the motion which can be extracted from the Noether's theorem considering variations
along the classical trajectories~\cite{dod_vol183,ocasta94}
	\begin{equation}
		A = \frac{i}{\sqrt{2}} \left( \varepsilon (t) \, p -
		\dot{\varepsilon} (t) \, q \right) \ ,
	\end{equation}
where the complex time-dependent function $\varepsilon (t)$ satisfy the classical
equation of motion
	\begin{equation}
		\ddot{\varepsilon} (t) + \omega^{2} (t) \, \varepsilon (t) = 0 \ ,
	\end{equation}
with initial conditions $\varepsilon (0) = 1$ and $\dot{\varepsilon} (0) = i$,
which led to satisfy the commutation relation
	\begin{equation}
		\left[ A , \, A^{\dagger} \right] = 1 \ .
	\end{equation}
We may find packet solutions of the Schr\"odinger equation which are
eigenstates of operator $A$ with complex eigenvalues $\alpha$. They have the form
	\begin{equation}
		\Psi_{\alpha} (q,t) = \Psi_{0} (q,t) \, \exp \left(-
		\frac{\left| \alpha \right|^{2}}{2} -
		\frac{\alpha^{2} \, \varepsilon^{\ast} (t)}{2 \, \varepsilon (t)} +
		\frac{\sqrt{2} \, \alpha \, q}{\varepsilon (t)} \right) \ ,
	\end{equation}
where
	\begin{equation}
		\Psi_{0} (q,t) = \pi^{-1/4} \frac{1}{\sqrt{\varepsilon (t)}} \,
		\exp \left( \frac{i \, \dot{\varepsilon} (t) \, q^{2}}{2 \,
		\varepsilon (t)} \right) \ .
	\end{equation}
Variances of the position and momentum of parametric oscillator in these
correlated coherent states can be calculated, and results are
	\begin{equation}
		\sigma_{q} = \frac{1}{2} \left| \varepsilon (t) \right|^{2} \ ,
		\qquad \sigma_{p} = \frac{1}{2} \left| \dot{\varepsilon} (t)
		\right|^{2} \ .
	\end{equation}
Thus, for $|\varepsilon (t)|<1$, the above states are squeezed states. The
correlation coefficient $r$ of the position and momentum has the
value corresponding to the minimum of the Robertson--Schr\"odinger
uncertainty relation
	\begin{equation}
		\sigma_{q} \, \sigma_{p} = \frac{1}{4(1 - r^{2})}
		\ .
	\end{equation}
Then for cases where $|\varepsilon (t)|^{2}$ or
$|\dot{\varepsilon} (t)|^{2}$ are less than 1 , the squeezing phenomenum
is present~\cite{elizabeth}.

Another example of squeezed states is related to the Caldirola--Kanai
Hamiltonian, which is used to describe dissipative systems. They
introduced a system described by the Lagrangian
	\begin{equation}
		L = f (t) \, \left\{ \frac{m}{2} \dot{q}^{2} - V (q)
		\right\} \ .
	\end{equation}
In this case, the equation of motion is
	\begin{equation}
		\ddot{q} + \frac{\dot{f}}{f} \dot{q} + \frac{1}{m}
		\frac{\partial V}{\partial q} = 0 \ .
	\end{equation}
The corresponding Hamiltonian is
	\begin{equation}
		H = \frac{1}{f (t)} \, \frac{p^{2}}{2 m} + f (t) \,
		V (q) \ ,
	\end{equation}
which can be quantized directly because it is quadratic in
position and momentum operators.  Generally $f (t) = \exp (2 \gamma t)$ is used
to be chosen.

In the same way as we did for the parametric oscillator, we can
define generalized annihilation and creation operators for the
quantum case through the relation
	\begin{equation}
		A (t) = \lambda_{q} q + \lambda_{p} p \ ,
	\end{equation}
where
	\begin{eqnarray}
		\lambda_{p} &&= \frac{1}{\sqrt{2 \sqrt{1-\gamma^{2}}}}
		\exp (- \gamma t) \left\{ i \, \exp (i \sqrt{1-\gamma^{2}} t)
		- \sin (\sqrt{1-\gamma^{2}} t) \right\} \ , \\
		\lambda_{q} &&= \frac{1}{\sqrt{2 \sqrt{1-\gamma^{2}}}}
		\exp (\gamma t) \big\{ (i \gamma + \sqrt{1-\gamma^{2}})
		 \, \exp (i \sqrt{1-\gamma^{2}} t) \\
		&& \qquad + \sqrt{1-\gamma^{2}}
		\, \cos (\sqrt{1-\gamma^{2}} t)
		- \gamma \, \sin (\sqrt{1-\gamma^{2}} t) \big\} \ .
	\end{eqnarray}
The generalized coherent wave functions are given by
	\begin{equation}
		\phi_{\alpha} (q,t) = (2\pi)^{-1/4} \frac{1}{\sqrt{\lambda_{p}}}
		\exp \left(-\frac{|\alpha|^{2}}{2}\right) \, \exp \left\{- \frac{i}{2 \lambda_{p}}
		\left( \lambda_{q} q^{2} - 2 \alpha q +
		i \alpha^{2} \lambda_{p}^{\ast} \right) \right\} \ .
	\end{equation}
The time-dependent probability distribution in position representation can
be rewritten as
	\begin{equation}
		\varrho_{\alpha} (q,t) = \frac{1}{\sqrt{2\pi} \, \sigma_{q} (t)}
		\exp \left\{ -\left(q - \frac{\langle q \rangle_{\alpha} (t) )^{2}}{
		(2 \sigma_{q}^{2} (t)} \right) \right\} \ .
	\end{equation}

The statistical properties of these states are given by
	\begin{eqnarray}
		&& \langle q \rangle_{\alpha} (t) = 2 \, \hbox{Im} (\lambda_{p}
		\alpha^{\ast}) \ , \qquad
		\langle p \rangle_{\alpha} (t) = 2 \, \hbox{Im} (\lambda_{q}^{\ast}
		\alpha) \ , \\
		&& \sigma_{q} (t) = \left| \lambda_{p} \right|^{2} \ , \qquad
		\sigma_{p} (t) = \left| \lambda_{q} \right|^{2} \ , \qquad
		\sigma_{pq} (t) = - \hbox{Re} (\lambda_{p} \lambda_{q}^{\ast})
		\ .
	\end{eqnarray}
The dispersion and correlation satisfy the Robertson--Schr\"odinger
uncertainty relation $\sigma_{p} \, \sigma_{q} -\sigma_{pq}^{2} =1/4$.

Evolution of the density probability function of
a coherent state under the action of the
Caldirola--Kanai Hamiltonian is shown in Fig. 1.


\section{Symplectic and optical tomograms}

The state of a quantum system is described by a Hermitian trace-class 
nonnegative density operator $\hat{\varrho}$. For a pure state, the density operator is a projector. For continuous variables (position or field quadrature), one can introduce the optical tomographic probability
distribution~\cite{ber-ber,vog-ris,raym}
    \begin{equation}
        {\cal W}_{\rm opt} (X,\theta) = \langle \delta \left(
        X - \cos\theta \, \hat{q} - \sin\theta \,
        \hat{p} \right) \rangle \ .
        \label{eq04}
    \end{equation}
This positive probability distribution, called optical
tomogram, is normalised for a normalised quantum state,
i.e.,
    \begin{equation}
        \int_{-\infty}^{\infty} dX \, {\cal W}_{\rm opt}
        (X,\theta) = 1 \ .
        \label{eq05}
    \end{equation}
It is important that the optical tomogram contains the same
information about the states as the density operator. The
optical tomogram determines completely the quantum state. It can
be considered as particular characteristics of the state analogous
to density matrix in position representation. The optical tomogram
can be extended to become the symplectic tomogram
    \begin{equation}
        {\cal W}_{\rm sym} (X,\mu,\nu) = \langle \delta \left(
        X - \mu \, \hat{q} - \nu \, \hat{p} \right)
        \rangle \ .
        \label{eq06}
    \end{equation}
Here $\mu$ and $\nu$ are real numbers. The random variable $X$ can
be treated as the position (field quadrature) measured in the
scaled and rotated reference frame of phase space. The symplectic
tomogram also determines the quantum state completely. It can be
used as characteristics of the state instead of density matrix in
position (or another) representation. The parameters $\lambda$ and
$\theta$, where
    \begin{equation}
        \mu = \be^{\lambda} \, \cos \theta \ ,
        \qquad \nu = \be^{-\lambda} \,
        \sin \theta \ ,
        \label{eq07}
    \end{equation}
describe scaling and rotation, respectively. Equation
(\ref{eq06}) can be rewritten using the expression for
the Wigner quasidistribution function of the state
$W(q,p)$~\cite{wig32}
    \begin{equation}
        {\cal W}_{\rm sym} (X,\mu,\nu) = \int
        \frac{dq \, dp}{2\pi} W(q,p) \, \delta \left(
        X - \mu \, q - \nu \, p \right) \ .
        \label{eq08}
    \end{equation}
This expression has the inverse
    \begin{equation}
        W(q,p) = \int \frac{dX \, d\mu \, d\nu}{2\pi}
        \be^{i \left( X - \mu \, q - \nu \, p \right)}
        \, {\cal W}_{\rm sym} (X,\mu,\nu) \ .
        \label{eq09}
    \end{equation}
The density operator $\hat{\varrho}$ can be expressed in terms of
the symplectic tomogram ${\cal W}_{\rm sym} (X,\mu,\nu)$
as follows \cite{d'ar96}
    \begin{equation}
        \hat{\varrho} = \frac{1}{2\pi} \int dX \,
        d\mu \, d\nu \ \be^{i \left( X - \mu \,
        \hat{q} - \nu \, \hat{p} \right)} \,
        {\cal W}_{\rm sym} (X,\mu,\nu) \ .
        \label{eq10}
    \end{equation}
The optical tomogram can be related to the Wigner function
    \begin{equation}
        {\cal W}_{\rm opt} (X,\theta) = \int
        \frac{dq \, dp}{2\pi} W (q,p) \, \delta
        \left( X - \cos\theta \, q - \sin\theta \,
        p \right) \ .
        \label{eq11}
    \end{equation}

\section{Squeeze tomograms}

We introduce another type of tomogram which we
call squeeze tomogram ${\cal W}_{\rm sq} (n,\mu,\nu)$. Here $n=0,
\, 1, \, 2, \, \dots$ has the physical meaning of the number of
photons in the quantum state of light under consideration. We
define the tomogram of the state with density operator
$\hat{\varrho}$ by the relation
    \begin{eqnarray}
        {\cal W}_{\rm sq} (n,\mu,\nu) &=&
        \langle n \vert \hat{\pmb{\cal S}}
        (\mu,\nu) \, \hat{\varrho} \,
        \hat{\pmb{\cal S}}^{\dagger} (\mu,\nu)
        \vert n \rangle \nonumber \\
        &=&
        \langle n \vert \hat{S} (\lambda)
        \, \hat{R} (\theta) \,
        \hat{\varrho} \,
        \hat{R}^{\dagger} (\theta) \,
        \hat{S}^{\dagger} (\lambda)
        \vert n \rangle \ .
        \label{eq17}
    \end{eqnarray}
Here $\hat{\pmb{\cal S}} (\mu,\nu) = \hat{S}
(\lambda) \, \hat{R} (\theta)$, where $\hat{S}
(\lambda)$ and $\hat{R} (\theta)$ are the squeezing
and rotation operators, respectively. They have
the form
    \begin{eqnarray}
        &&\hat{S} (\lambda) = \exp \left[
        \frac{i \lambda}{2} \left( \hat{q}
        \hat{p} + \hat{p} \hat{q} \right)
        \right] \ ,
        \label{eq18} \\
        &&\hat{R} (\theta) = \exp \left[
        \frac{i \theta}{2} \left(
        \hat{q}^{2} + \hat{p}^{2} \right)
        \right] \ .
        \label{eq19}
    \end{eqnarray}
The scaling parameter $\lambda$ and rotation
angle $\theta$ are connected with symplectic
transform parameters $\mu$ and $\nu$
by~Eq.~(\ref{eq07}).

The squeeze tomogram can be interpreted as
the diagonal matrix element in a Fock basis of
the scaled and rotated density operator
    \begin{equation}
        \hat{\varrho}^{\mu \nu} =
        \hat{\pmb{\cal S}} (\mu,\nu) \,
        \hat{\varrho} \,
        \hat{\pmb{\cal S}}^{\dagger} (\mu,\nu)
        \ .
        \label{eq20}
    \end{equation}
Since the squeezing and rotation are unitary operators, the
Hermitian nonnegative density operator $\hat{\varrho}^{\mu \nu}$
has positive diagonal matrix elements in the Fock basis. These
matrix elements (tomograms) have the physical meaning of photon
distribution functions in the state described by the density
operator $\hat{\varrho}^{\mu \nu}$. To measure the tomogram one
has to take the initial photon state with density operator
$\hat{\varrho}$. Then one needs to rotate the quadratures as it is
done in the homodyne detection scheme. The rotated state has to be
squeezed by applying the squeezing operator $\hat{S}^{\dagger}
(\lambda)$. Measuring the photon statistics in the obtained state
with density operator $\hat{\varrho}^{\mu\nu}$ one gets the
squeeze tomogram ${\cal W}_{\rm sq} (n,\mu,\nu)$. This tomogram is
the normalized probability distribution of the discrete random
variable $n$. The tomogram is normalized, satisfying the equality
    \begin{equation}
        \sum_{n=0}^{\infty}
        {\cal W}_{\rm sq} (n,\mu,\nu) = 1 \ .
        \label{eq21}
    \end{equation}
The tomogram depends on the number of
photons $n$ and two real parameters $\mu$
and $\nu$. The number of the parameters is
sufficient to characterize the quantum
state completely, since it is determined by
the Wigner function depending on two real
variables $q$ and $p$.

Let us find out the connection of the introduced
squeeze tomogram with other characteristics of
photon quantum states, e.g., with density operator
(density matrix) in position representation. This
connection can be presented in the form of
integral transform of the density matrix
    \begin{eqnarray}
        &&{\cal W}_{\rm sq} (n,\mu,\nu) \equiv
        {\cal W}_{\rm sq} (n,\lambda,\theta) \nonumber \\
        && \quad = \int dx \, dy \, \varrho (x,y)
        \ {\cal K} (x,y,n,\mu,\nu) \ .
        \label{eq22}
    \end{eqnarray}
The kernel of the integral transform has the form
    \begin{eqnarray}
        &&{\cal K} (x,y,n,\mu,\nu) = \frac{1}{\sqrt{\pi
        (\mu^{2}+\nu^{2})} 2^{n} n!} \nonumber \\
        && \ \times H_{n} \left( \frac{x}{\sqrt{\mu^{2}
        +\nu^{2}}} \right) \ H_{n} \left(
        \frac{y}{\sqrt{\mu^{2}+\nu^{2}}} \right)
        \nonumber \\
        && \ \times \exp \left\{ -i \frac{x^{2}}{2}
        \left[ \frac{\sqrt{2}}{1-\sqrt{1-4 \mu^{2}
        \nu^{2}}} - \frac{\mu+i\nu}{\nu (\mu^{2}+\nu^{2})}
        \right] \nonumber \right. \\
        && \left. \quad + i \frac{y^{2}}{2} \left[
        \frac{\sqrt{2}}{1-\sqrt{1-4 \mu^{2} \nu^{2}}} -
        \frac{\mu-i\nu}{\nu (\mu^{2}+\nu^{2})} \right]
        \right\} \ ,
        \label{eq23}
    \end{eqnarray}
where $H_{n}$ denotes the Hermite polynomial of order $n$.
The derivation of this formula is given in \cite{JPAlast}.

One can find the relation of squeeze tomogram to the Wigner function. The connection of squeeze tomogram with the Wigner function can be presented
in the integral form
    \begin{equation}
        {\cal W}_{\rm sq} (n,\mu,\nu) =
        \int dq \, dp \, W (q,p) \,
        {\cal K}_{W} (q,p,n,\mu,\nu) \ .
        \label{eq24}
    \end{equation}
The kernel of the integral transform has the form
    \begin{eqnarray}
        && {\cal K}_{W} (q,p,n,\mu,\nu)
        \nonumber \\
        && \qquad = \frac{(-1)^{n}}{\pi}\, \exp
        \left( - \left| z \right|^{2} /2 \right)
        \, L_{n} \left( \left| z \right|^{2}
        \right) \ ,
        \label{eq25}
    \end{eqnarray}
with
    \begin{eqnarray}
        && \left| z \right|^{2} = \frac{2 q^{2}}{\mu^{2}
        +\nu^{2}} + 2 \left( \mu^{2}+\nu^{2} \right)
        \nonumber \\
        &&\times\left[p \quad - \left( \frac{\sqrt{2}}{1
        -\sqrt{1-4 \mu^{2} \nu^{2}}} - \frac{\mu}{\nu(
        \mu^{2}+\nu^{2})} \right) \, q \right]^{2} \ .
        \label{eq26}
    \end{eqnarray}
For $\nu=1$, $\mu=0$ (or $\theta=\pi/2$, $\lambda=0$),
which means that there is no squeezing and a $\pi/2$
rotation, the obtained kernel coincides with the
Wigner function of the Fock state~(given
in~\cite{dod_vol183}).

The symplectic tomograms can be written within the framework of
the star-product quantization~\cite{ovmanko}. Then it is
associated with the set of operators
    \begin{eqnarray}
        \hat{U} (\vec{x}) &=& \delta
        \left( X - \hat{\pmb{\cal S}}^{\dagger} (\mu,\nu)
        \, \hat{q} \, \hat{\pmb{\cal S}} (\mu,\nu) \right)
        \ ,
        \label{eqnew01} \\
        \hat{D} (\vec{x}) &=& \frac{1}{2\pi}
        \exp \left\{ i \left( X - \mu \hat{q} - \nu \hat{p}
        \right) \right\} \ ,
        \label{eqnew02}
    \end{eqnarray}
with $\vec{x} = (X,\mu,\nu)$. According to the star-product quantization scheme, the symplectic tomogram is the tomographic symbol of the density operator and it is given by
    \[
        {\cal W} (\vec{x}) = f_{\hat{\varrho}}
        (\vec{x}) =
        \hbox{Tr} \left\{ \hat{\varrho} \,
        \hat{U} (\vec{x}) \right\} \ .
    \]
The density operator is expressed in terms of symplectic tomogram
    \[
        \hat{\varrho} = \int dX \, d\mu \, d\nu \,
        {\cal W}_{\hat{\varrho}} (X,\mu,\nu) \,
        \hat{D} (X,\mu,\nu) \ .
    \]
Suppose that one uses the other star-product scheme described by
the vector $\vec{y} = (n,\mu^{\prime},\nu^{\prime})$ and sets of
operators $\hat{U}^{\prime} (\vec{y})$ and $\hat{D}^{\prime}
(\vec{y})$. Thus we introduce the squeeze tomogram as another tomographic
symbol of the density operator
    \[
        {\cal W} (\vec{y}) =
        \phi_{\hat{\varrho}} (\vec{y}) =
        \hbox{Tr} \left\{ \hat{\varrho} \,
        \hat{U}^{\prime} (\vec{y}) \right\} \ .
    \]
The inverse relation reads
    \[
        \hat{\varrho} = \sum_{n} \int d\mu^{\prime}
        \, d\nu^{\prime} \, {\cal W}_{\hat{\varrho}}
        (n,\mu^{\prime},\nu^{\prime}) \,
        \hat{D}^{\prime}
        (n,\mu^{\prime},\nu^{\prime}) \ .
    \]
The operator $\hat{U}^{\prime}
(n,\mu^{\prime},\nu^{\prime})$ is given by the expression
    \begin{equation}
        \hat{U}^{\prime}
        (n,\mu^{\prime},\nu^{\prime}) = \delta \left(
        n - \hat{\pmb{\cal S}}^{\dagger}
        (\mu^{\prime},\nu^{\prime}) \hat{a}^{\dagger}
        \hat{a} \, \hat{\pmb{\cal S}}
        (\mu^{\prime},\nu^{\prime}) \right) \ ,
        \label{eqnew03}
    \end{equation}
where $\hat{a}^{\dagger}$ and $\hat{a}$ are boson creation and
annihilation operators.

Two different symbols of the same operator can be related
through the expression
    \[
        \phi_{A} (\vec{y}) = \int d\vec{x} \, f_{A}
        (\vec{x}) \, \hbox{Tr} \left\{ \hat{D}
        (\vec{x}) \, \hat{U}^{\prime} (\vec{y})
        \right\} \ .
    \]
Therefore
    \begin{eqnarray}
        && {\cal W}_{\rm sq} (n,\mu^{\prime},\nu^{\prime}) =
        \int dX \, d\mu \, d\nu \ {\cal W}_{\rm sym} (X,\mu,\nu)
        \nonumber \\
        && \quad \times {\cal K}_{S}
        (n,\mu^{\prime},\nu^{\prime}, X,\mu,\nu).
        \label{eqnew04}
    \end{eqnarray}
The kernel has the form~\cite{JPAlast}
    \begin{equation}
        {\cal K}_{\cal S}
        (n,\mu^{\prime},\nu^{\prime},X,\mu,\nu) =
        \frac{\be^{i X}}{2\pi}
        \be^{-\left| \alpha \right|^{2}/2}
        \, L_{n} \left( \left| \alpha \right|^{2}
        \right) \ .
        \label{eq28}
    \end{equation}
Here $L_{n}$ is a Laguerre polynomial and the complex
variable $\alpha$ reads
    \begin{equation}
        \alpha = \frac{1}{\sqrt{2}} \left(
        \tilde{\nu} - i \tilde{\mu} \right) \ ,
        \label{eq29}
    \end{equation}
where $\tilde{\mu}$ and $\tilde{\nu}$ are given by
\begin{eqnarray}
        \tilde{\mu} &=& -\frac{\nu^{\prime}}{2 \nu}
        \left( 1 - \sqrt{1 - 4 \mu^{2} \nu^{2}} \right) +
        \mu^{\prime} \mu \ ,
        \label{eqnew09} \\
        \tilde{\nu} &=& \frac{\nu^{\prime}}{2 \mu}
        \left( 1 + \sqrt{1 - 4 \mu^{2} \nu^{2}} \right) +
        \mu^{\prime} \nu \ .
        \label{eqnew10}
    \end{eqnarray}

\section{Examples}

In this section we consider several examples of
squeeze tomograms of important states of photons.
The first example is the ground or vacuum state of
the electromagnetic field with density operator
    \begin{equation}
        \hat{\varrho}_{v} = \vert 0 \rangle
        \langle 0 \vert \ .
        \label{ex01}
    \end{equation}
The squeeze tomogram of this state reads
    \begin{equation}
        {\cal W}_{0} (n,\lambda,\theta) =
        \left| \langle n \vert \hat{S}
        (\lambda) \vert 0 \rangle \right|^{2}
        \ ,
        \label{ex02}
    \end{equation}
where $\hat{S} (\lambda)$ is the unitary squeezing
operator. The tomogram in explicit form reads
    \begin{equation}
        {\cal W}_{0} (n,\lambda,\theta) =
        \frac{\left( - \tanh \lambda \right)^{n}}
        {n! \, 2^{n} \, \cosh \lambda} \, \left\{
        H_{n} (0) \right\}^{2} \ .
        \label{ex03}
    \end{equation}
One can see that the angle $\theta$ is not present
in the tomogram for the vacuum state.


Another important state is the coherent state
$\vert \alpha \rangle$ of the photon with the
density operator
    \begin{equation}
        \hat{\varrho}_{\alpha} = \vert \alpha
        \rangle \langle \alpha \vert \ .
        \label{ex04}
    \end{equation}
According to definition, the squeeze tomogram for
this state reads
    \begin{equation}
        {\cal W}_{\alpha} (n,\lambda,\theta)
        = \left| \langle n \vert \hat{S}
        (\lambda) \hat{R} (\theta) \vert \alpha
        \rangle \right|^{2} \ .
        \label{ex05}
    \end{equation}
One can easily show that
    \[
        \langle n \vert \hat{S} (\lambda)
        \hat{R} (\theta) \vert \alpha \rangle
        = \be^{(\lambda + i \theta)/2} \, \int
        dx \, \psi_{n}^{\ast} (x) \,
        \psi_{\tilde{\alpha}} (\be^{\lambda} x)
        \ ,
    \]
with $\tilde{\alpha} = \alpha \be^{i \theta}$. We can get
    \begin{eqnarray*}
        && \sum_{n=0}^{\infty} \frac{\beta^{\ast \, n}}
        {\sqrt{n!}} \langle n \vert \hat{S} (\lambda)
        \hat{R} (\theta) \vert \alpha \rangle \\
        && \ = \frac{\be^{i \theta/2}}{\sqrt{\cosh
        \lambda}} \, \exp \left\{ -\frac{\left| \alpha
        \right|^{2}}{2} + \frac{1}{2} \tilde{\alpha}^{2}
        \tanh \lambda \right\} \\
        && \quad \times \exp \left\{ -\frac{1}{2}
        \beta^{\ast \, 2} \tanh \lambda +
        \frac{\tilde{\alpha}}{\cosh \lambda} \beta^{\ast}
        \right\} \ .
    \end{eqnarray*}
By means of the generating function of the Hermite polynomials
we obtain the matrix element
    \begin{eqnarray}
        && \langle n \vert \hat{S} (\lambda)
        \hat{R} (\theta) \vert \alpha \rangle
        = \be^{(-\left| \alpha \right|^{2}
        + i\theta +
        (\alpha \be^{i \theta})^{2}
        \tanh\lambda) /2} \nonumber \\
        && \quad \times \sqrt{\frac{\left| \tanh \lambda
        \right|^{n}} {2^{n} n! \cosh \lambda}}
        H_{n} \left( \frac{\alpha \be^{i\theta}}
        {\sqrt{\left| \sinh 2 \lambda \right|}}
        \right) \ .
        \label{ex06}
    \end{eqnarray}

For $\alpha=0$, if we take the absolute value of the last
expression, we get Eq.~(\ref{ex03}). One can see
that the tomograms~(\ref{ex03}) and~(\ref{ex05})
coincide with the photon distribution function of
squeezed vacuum and generic squeezed coherent states,
respectively. These photon distributions are given,
e.g., in~\cite{dod_vol183}. In Fig.~2
we illustrate the behaviour of the tomogram of a
coherent state as a function of $n$ and $\lambda$,
using $\alpha=3$ and $\theta=0$.

Another specific example is the Fock state of the photon
$\vert m \rangle$ with density operator
    \begin{equation}
        \hat{\varrho}_{m} = \vert m \rangle
        \langle m \vert \ , \qquad m = 0, 1,
        \dots \ .
        \label{ex11}
    \end{equation}
The squeeze tomogram of this state is
    \begin{equation}
        {\cal W}_{m} (n,\lambda) =
        \left| \langle n \vert \hat{S} (\lambda)
        \vert m \rangle \right|^{2} \ .
        \label{ex12}
    \end{equation}
In fact, it is modulus squared of the matrix
element of the squeezing operator in Fock basis.
It does not depend on rotation angle $\theta$.
Again, the squeeze tomogram coincides with the
photon distribution function of the squeezed Fock
state. The example of the tomogram for the Fock
state $\vert 1 \rangle$ reads
    \begin{equation}
        {\cal W}_{1} (n,\lambda) =
        \frac{n^{2}}{2^{n-1} n!}
        \frac{(\tanh \lambda)^{n-1}}
        {(\cosh \lambda)^{3}} \,
        \left[ H_{n-1} (0) \right]^{2}
        \ .
        \label{ex14}
    \end{equation}

The even and odd coherent states (Schr\"odinger
cat states)~\cite{dmm74} are paradigmatic
examples of superposition of quantum states.
The density operators for these states read
    \begin{equation}
        \hat{\varrho}_{\alpha}^{\pm} =
        \left| {\cal N}_{\pm} \right|^{2}
        \left( \vert \alpha \rangle \pm
        \vert -\alpha \rangle \right)
        \left( \langle \alpha \vert \pm
        \langle -\alpha \vert \right) \ ,
        \label{ex15}
    \end{equation}
where
    \begin{equation}
        {\cal N}_{\pm} = \sqrt{\frac{1}
        {2 \left( 1 \pm \be^{-2 \left|
        \alpha \right|^{2}}\right)}}
        \ .
        \label{ex16}
    \end{equation}
The squeeze tomograms for the Schr\"odinger
cat states are
    \begin{eqnarray}
        && {\cal W}_{\alpha}^{\pm}
        (n,\lambda,\theta) = \frac{1}
        {1 \pm \be^{-2 \left|
        \alpha \right|^{2}}} \Big\{
        \left| \langle n \vert \hat{S} (\lambda)
        \hat{R} (\theta) \vert \alpha \rangle
        \right|^{2} \nonumber \\
        && \qquad \pm \hbox{Re} \Big[
        \langle n \vert \hat{S} (\lambda)
        \hat{R} (\theta) \vert \alpha \rangle
        \langle n \vert \hat{S} (\lambda)
        \hat{R} (\theta)
        \vert -\alpha \rangle^{\ast} \Big]
        \Big\} \nonumber \\
        && \ = \frac{1} {1 \pm \be^{-2 \left|
        \alpha \right|^{2}}} \left[ 1 \pm
        (-1)^{n} \right] \, {\cal W}_{\alpha}
        (n,\lambda,\theta) \ .
        \label{ex17}
    \end{eqnarray}
We observe terms which are due to interference of
states $\vert \alpha\rangle$ and $\vert -\alpha\rangle$.

The example of a mixed state tomogram (thermal state of light)
with density operator
    \begin{eqnarray}
        &&\hat{\varrho}_{T} = \frac{1}{Z}
        \be^{-\frac{1}{T} \left(
        \hat{a}^{\dagger} \hat{a} + 1/2
        \right)} \ , \nonumber \\
        &&Z = \sum_{n=0}^{\infty}
        \be^{-\frac{1}{T}(n+1/2)}
        = \frac{1}{2} \, \hbox{cosech} \left(
        \frac{1}{2T} \right)
        \ ,
        \label{ex18}
    \end{eqnarray}
is given by the sum
    \begin{eqnarray}
        && {\cal W}_{T} (n,\lambda,\theta)
        = \frac{1}{Z} \sum_{m=0}^{\infty}
        \be^{-\frac{1}{T}(m+1/2)}
        {\cal W}_{m} (n,\lambda)
        \nonumber \\
        && \ = \frac{1}{Z} \sum_{m=0}^{\infty}
        \be^{-\frac{1}{T}(m+1/2)}
        \frac{\hbox{sech}~\lambda}{m!n!}
        \left[ H^{\{ {\cal R}\}}_{n \, m} (0)
        \right]^{2} \ .
        \label{ex19}
    \end{eqnarray}
The matrix $\cal R$ is given by
    \[
        {\cal R} = \left( \begin{array}{cc}
        \tanh~\lambda & -\hbox{sech}~\lambda \\
        -\hbox{sech}~\lambda & -\tanh~\lambda
        \end{array} \right) \ .
    \]
One can see that the squeeze tomogram does not
depend on the rotation angle $\theta$ because
it contains the sum of Fock state tomograms,
and these tomograms do not depend on the
rotation angle. For $T\rightarrow 0$, the
tomogram is going to the tomogram of the
vacuum state.

\section{Conclusions}

We discussed different aspects of squeezing operator considered by Pleba\'nsky long ago~\cite{Pleb54,InfPleb,Pleb55,Pleb} and its application to the problem of nonclassical (squeezed) states and to the problem of squeeze tomography.
We have shown that for systems with time-dependent Hamiltonians, e.g., for oscillator with time-dependent frequency and for damped ocsillator, the squeezed states appear naturally in the process of time evolution. We constructed also the squeeze tomogram which is a fair probability distribution of discrete random variable. The squeeze tomogram depends on extra two parameters and it describes the quantum state. The squeeze tomogram is related to other functions depending the quantum state including the Wigner function, symplectic and optical tomograms by means of integral transform. The squeeze tomography can be discussed as complimentary method to measure quantum states of photon.

\section*{Acknowledgments}

M.A.M. and V.I.M. thank Institute of Nuclear Sciences of UNAM and Organizers of the Conference ``Topics in Mathematical Physics, General Relativity and Cosmology on the Ocassion of the 75th Birthday of Professor Jerzy F. Pleba\'nski'' for kind hospitality. M.A.M. thanks the Russian Foundation for Basic Research for partial support under Project~no.~03-02-16408.

\newpage

\section*{Figure captions}

{\bf Figure 1.} Evolution of the density probability function of
a coherent state with amplitude $\alpha = 1/2$ under the action of the
Caldirola--Kanai Hamiltonian with $\gamma = 0.1$. In the lower part of
the figure, there is a contour density plot where the phenomenon of
squeezing is evident when time increases.

\

\noindent
{\bf Figure 2.} Squeeze tomogram of a coherent
state. We used the parameters $\alpha=3$ and $\theta=0$,
which characterize a nonrotated phase space frame.

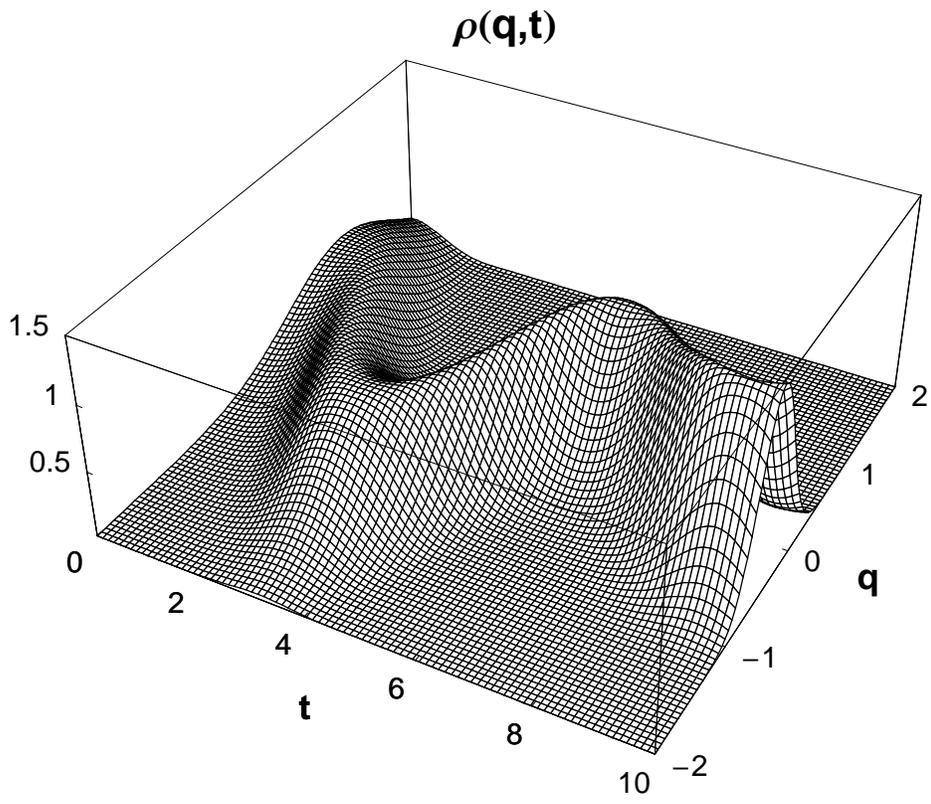
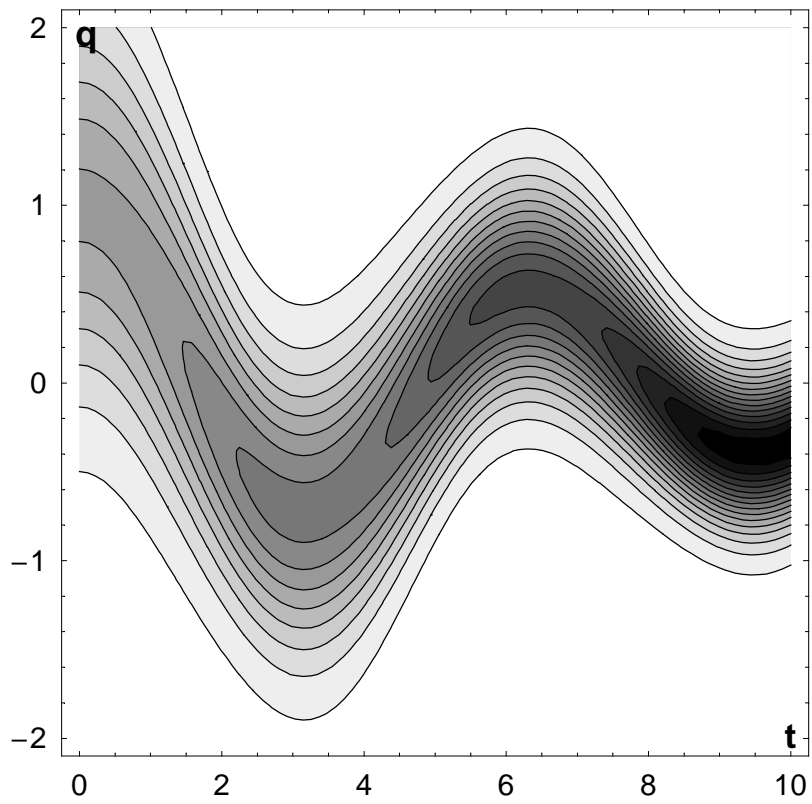
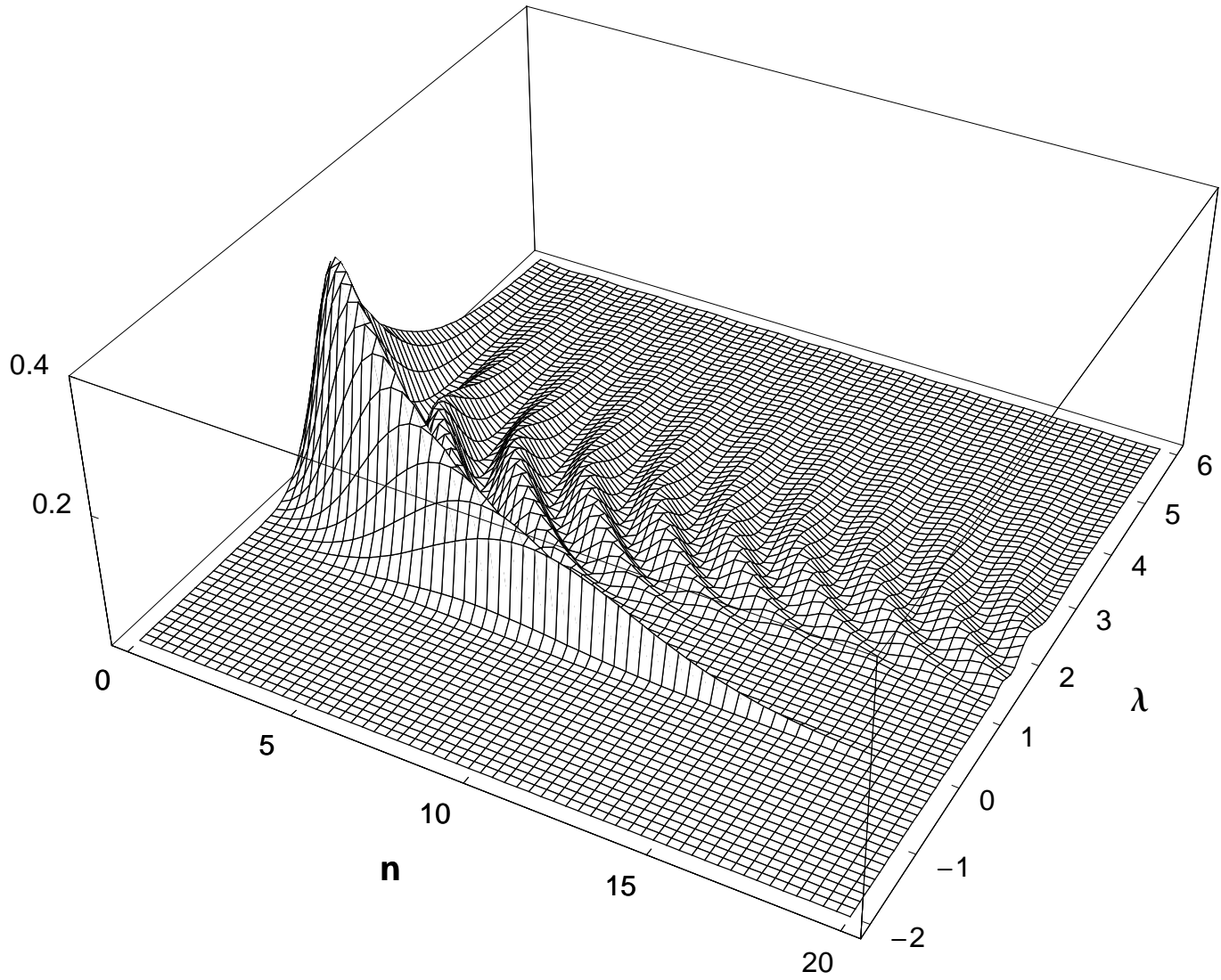
\end{document}